\documentclass[graybox]{svmult}

\usepackage{geometry}
\usepackage{ebgaramond,ebgaramond-maths,makeidx,graphicx,multicol}
\usepackage[bottom]{footmisc}

\begin{document}
\title*{A temporal network version of Watts's cascade model}
\author{Fariba Karimi and Petter Holme}
\institute{Fariba Karimi \at IceLab, Department of Physics, Ume{\aa} University, 90187 Ume\aa, Sweden \and
Petter Holme \at IceLab, Department of Physics, Ume{\aa} University, 90187 Ume\aa, Sweden\\ \email{petter.holme@physics.umu.se}\newline 
Department of Energy Science, Sungkyunkwan University, Suwon 440--746, Korea\\ \email{holme@skku.edu}\newline Department of Sociology, Stockholm University, 10691 Stockholm, Sweden}
\maketitle

\abstract*{Threshold models of cascades in the social sciences and economics explain the spread of opinion and innovation due to social influence. In threshold cascade models, fads or innovations spread between agents as determined by their interactions with other agents and their personal threshold of resistance. Typically, these models do not account for structure in the timing of interaction between the units. In this work, we extend a model of social cascades by Duncan Watts to temporal interaction networks. In our model, we assume friends and acquaintances influence agents for a certain time into the future. That is the influence of the past ages and becomes unimportant. Thus, our modified cascade model has an effective time window of influence. We explore two types of thresholds---thresholds to fractions of the neighbors or absolute numbers. We try our model on six empirical datasets and compare them with null models.}

\abstract{Threshold models of cascades in the social sciences and economics explain the spread of opinion and innovation due to social influence. In threshold cascade models, fads or innovations spread between agents as determined by their interactions with other agents and their personal threshold of resistance. Typically, these models do not account for structure in the timing of interaction between the units. In this work, we extend a model of social cascades by Duncan Watts to temporal interaction networks. In our model, we assume friends and acquaintances influence agents for a certain time into the future. That is the influence of the past ages and becomes unimportant. Thus, our modified cascade model has an effective time window of influence. We explore two types of thresholds---thresholds to fractions of the neighbors or absolute numbers. We try our model on six empirical datasets and compare them with null models.}

\section{Introduction}
\label{sec:1}

Threshold models of cascades have been studied extensively in the social-science literature. The examples of the phenomena they seek to explain include diffusion of innovation, rumors, diseases, strikes, voting behavior, and migration. Diffusion models such as Bass model~\cite{bass_diffusion_innovasion}  are canonical models. These assume that the individual choice of action is independent of the individuals' total number of interactions. However, sociologist Mark Granovetter~\cite{granovetter_threshold} proposed thresholds to influence as a rational action. This insight comes from that there must be a point where the net benefits of acting exceed the net costs. This also reflects the fact that many decisions need group size to be considered. For example, the costs of joining a riot would probably decrease as the group size increases. Granovetter further proposes two main factors that influence spreading in threshold scenarios---social structure and timing of social action. In this chapter, we focus on the latter factor.

Social psychologist Bibb Latan\'e~\cite{latane_social_impact} proposed social impact theory to explain social influence as a multiplicative function of strength, immediacy (an inverse function of physical distance), and the number of sources of influence.  In the threshold model of cascades, we only investigate the effects of the third factor (that the influence is proportional to the number of people involved). The threshold itself corresponds to the individual's persistence to change his or her state. Threshold models have been studied for uniform or heterogeneous threshold values~\cite{valente_diffusion,granovetter_threshold}. Dodds and Watts~\cite{dodds_contagion_2004} showed that minor manipulation of individual thresholds could have a major impact on spreading phenomena. Here, for simplicity (like in Ref.~\cite{watts_threshold}), we assume an identical threshold value for all individuals.

This chapter is inspired by Watts' threshold model of cascades in networks~\cite{watts_threshold}. This work explains, for example, how innovation can trigger a cascade of adaptation solely by network interactions. Even though this approach does not represent the reality of innovation adaptation, it estimates the importance of networks in the spreading process. In threshold models, the agents' decision of action is a binary choice, such as replacing an old production method with a new one. In this model, the individuals' choice of action depends only on other neighbors' choices. In Schelling's words, we have a model of binary choice with externalities.

We extend the work of Ref.~\cite{karimi_holme} where we present a generalization of a threshold model to temporal networks~\cite{holme_temp_ntw,karsai_slow}. Temporal networks, the theme of this book, are networks that encode information about when things happen, not only about which nodes are in contact. We contrast our generalized model to the behavior of the original, static-network version.  In our model (as in Watts's model), all the agents initially have the same state, and the thresholds are homogeneously distributed over the population. During the simulation, the agents change their states according to their interaction with their neighborhood and the time of these interactions. We study two types of thresholds---fractional and absolute---corresponding to whether the individual responds to the fraction of neighbors with a deviating opinion or the absolute number of such neighbors. We note that while fractional thresholds are most common in the social and economic literature, absolute threshold models are used in bootstrap percolations and self-organized criticality, which focus on local dependency~\cite{balough_bootstrap,fontes_bootstrap}.

\section{Methods}
\label{sec:2}

We consider a system of \textit{vertices} (agents) and \textit{edges} (vertices that at some point are in contact with each other). The system can be represented as a network $G$ with a set of vertices $V$ and a set of edges $E$. Every edge is associated with a set of times of interaction events. These interactions are bidirectional.

Each vertex or agent in the network has a \textit{state}. The state can be either 0 or 1. We call vertices with state 0 \textit{non-adopters}. Vertices with state 1 correspond to \textit{adopters}. Initially, all the vertices in the network have state 0 (non-adopters). We start by randomly choosing a vertex and assign the state to 1. This initial vertex with state 1 can, according to the diffusion-of-innovation literature~\cite{valente_diffusion}, be interpreted as an innovator advertising a new product. Here, for simplicity, we only divide the population into adopters and non-adopters. We thus study cascades of adoption from adopters to non-adopters. Choosing random vertex and repeating the simulation for many realizations ensures that the results come from large-scale network properties rather than the initial agent's position in the network.

We run the dynamic of interactions in chronological order. Influential interactions are those that occur within a \textit{time window} $\theta$ in the past. This means agents change their initial state 0 to 1, depending on the contacts within the mentioned time window. If the fraction (or number) of adopters among the agent's neighbors exceeds the threshold value within the time window, then the agent changes state to 1. In our (and Watts') setup, an adopter remains an adopter for the rest of the simulation. The motivation of the latter is to simplify the model.

Let us denote the total population by $N$ and the number of adopters (as a function of time) $N_\mathrm{adopters}(t)$. The cascade size is calculated as follows (and averaged over at least 200 runs of simulations)
\begin{equation}\label{fig:adopters}
\Omega (t) = \frac{\langle N_\mathrm{adopters}(t)\rangle}{N}
\end{equation}
Let $\phi$ be the threshold value in which agent change their state accordingly.  Let $a_{i}$ denote the number of adopters vertex $i$ meets in a time window $\theta$. Here $c_{i}$ denotes total number of contacting neighbors that vertex $i$ has within the time window $\theta$. In the case of fractional threshold values, we count the fraction of adopters in the neighborhood of a vertex $i$ within a time window as follows
\begin{equation}
f_{i}=\frac{a_{i}}{c_{i}}
\end{equation}
The state of adoption occurs according to the following conditions
\begin{equation}\label{eq:fractional_threshold}
Pr[\mbox{adopting from vertex $i$}] =\left\{\begin{array}{ll} 1 & \mbox{if~}  f_{i}  \geq \phi \\ 0 & \mbox{if~}  f_{i} < \phi \end{array}\right .
\end{equation}

For the case of absolute threshold model, let $\Phi$ be the absolute number of adopters that vertex $i$ needs to meet within the time window $\theta$ to be able to change its state to 1. The condition for the change of state is the following
\begin{equation}\label{eq:absolute_threshold}
Pr[\mbox{adopting from vertex $i$}] =\left\{ \begin{array}{ll} 1 &   \mbox{if~}  f_{i} c_{i} \geq \Phi \\ 0 & \mbox{if~} f_{i}c_{i} < \Phi \end{array}\right .
\end{equation}

\begin{figure}
  \includegraphics[width=\linewidth]{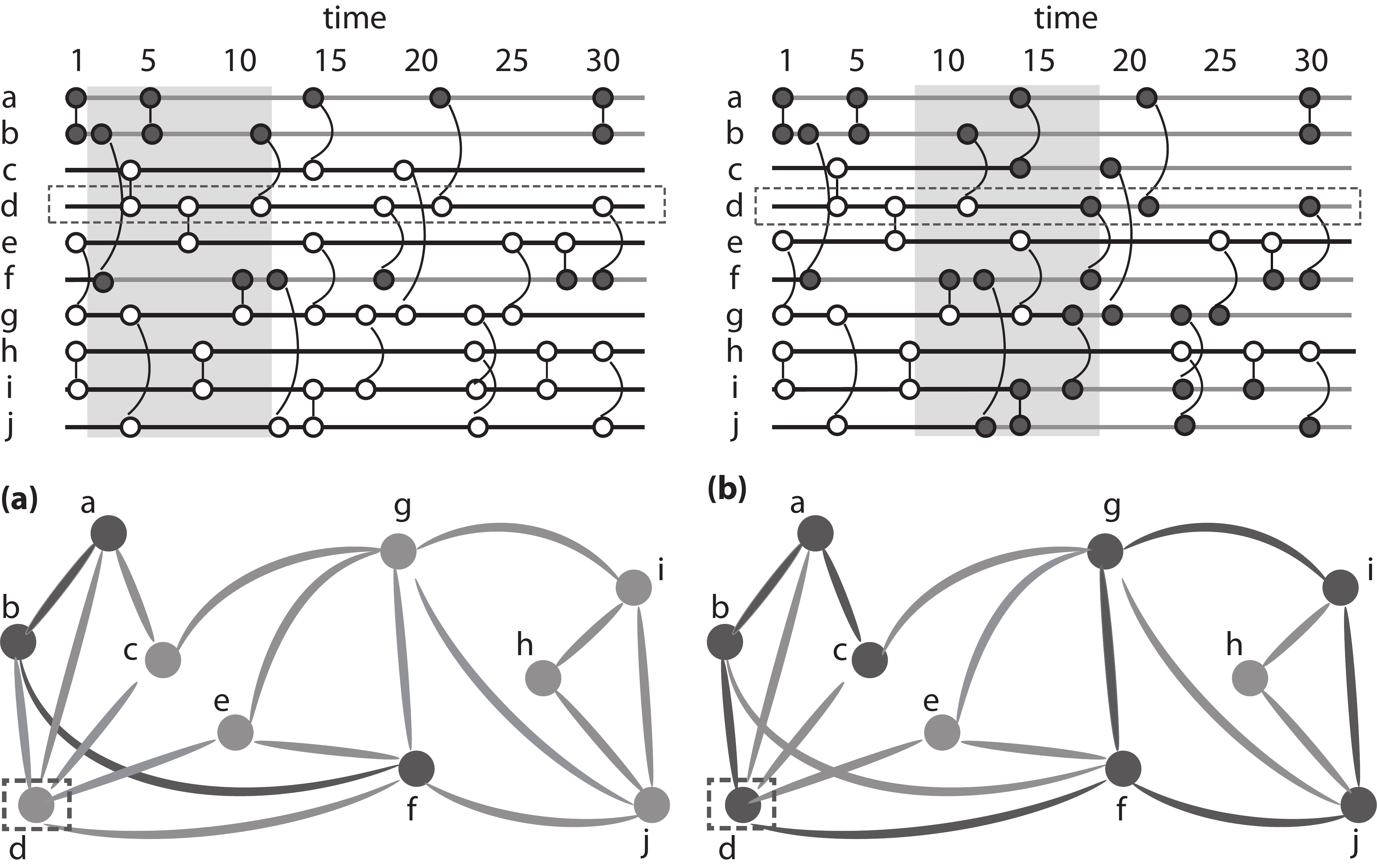}
  \caption{An illustration of our temporal-network
cascade model. In the panel (a), vertex \emph{d} does not change its state
according to the time window and value of threshold. In panel (b),
vertex \emph{d} reaches the criteria to change its state.}\label{fig:cascade-illustration}
\end{figure}

Fig.~\ref{fig:cascade-illustration} illustrates the model. Contacts are illustrated in a timeline. Shaded circles indicate adopters, white or gray circles indicate non-adopters. The threshold here is assumed to be $\phi=0.5$ and the time window $\theta=10$. As time progresses,  the time window slides through each contact, and the vertices are updated according to their contacts within the time window. In panel (a), vertex d does not change its state even though it is in contact with an adopter. At another time, panel (b), vertex d changes its color because the adopters' fraction exceeds the threshold inside the current time window.

\section{Empirical datasets}
\label{sec:3}
We test our model on six empirical datasets stemming from different
types of human interactions. The datasets were obtained with all individuals anonymized to protect their privacy. All datasets have been used in previous studies. The first dataset consists of self-reported
sexual contacts from a Brazilian online forum where the sex-buyers rate
and discuss female sex-sellers~\cite{rocha_epidemic}. The second dataset comes from an email exchange at a university~\cite{ekman_email}. Ref.~\cite{barabasi_burst_2005} used it to argue that human behavior often comes in bursts. The third dataset was collected at a three-day conference from face-to-face interactions between conference attendees~\cite{ht_contact_data}. The fourth dataset comes from a Swedish Internet dating site where the interaction ranges from partner seeking to friendship oriented~\cite{puss_och_kram}. The fifth and sixth datasets come from a Swedish forum for rating and
discussing films~\cite{filmtipset_geman}. One of these datasets represents comments in a forum organized to see who comments on whom. The other datasets come from email-like messages. Table~\ref{tab:Properties-of-datasets} summarizes details of datasets such as number of vertices, number of contacts, sampling time, and time resolution. Some of the datasets like movie forums, email, and conference contacts can be an underlying structure for spreading social influence, like fads or ideas or computer viruses. The sexual-contact dataset and the online dating datasets represent structures over which sexually transmitted infections can potentially spread.

\begin{table*}
\caption{\label{tab:Properties-of-datasets}Summary of properties of the datasets.}
\begin{tabular}{c|cccc}
Data & No.\ vertices & No.\ contacts & time duration & resolution\tabularnewline \hline
Email & 318,8 & 309,125 & 82 days & second\tabularnewline
Online dating & 293,41 & 536,276  & 512 days & second\tabularnewline
Internet community messages & 329 & 434  & 500 days & second\tabularnewline
Internet community forum  & 2384 & 291,151  & 500 days & second\tabularnewline
Prostitution & 167,30 & 506,32 & 223,2 days & day\tabularnewline
Conference &  113 & 20,818 & 3 days & 20 seconds\tabularnewline
\end{tabular}
\end{table*}

\section{Fractional-threshold model}
\label{sec:4}
\subsection{Effect of threshold values and time windows on cascade size}
\label{subsec:1}

In this section, we investigate the effect of varying time windows  $\theta$ and threshold values $\phi$ on the size of cascade. The cascade size $\omega$ is defined as the fraction of adopters over the whole population. The size of cascade can vary from 0 to 1. Results in all sections are the outcome of at least 200 simulation runs with different random seed. Therefore, the size of cascade we report is the average over that many simulation runs.

\begin{figure*}
  \includegraphics[width=1.0\linewidth]{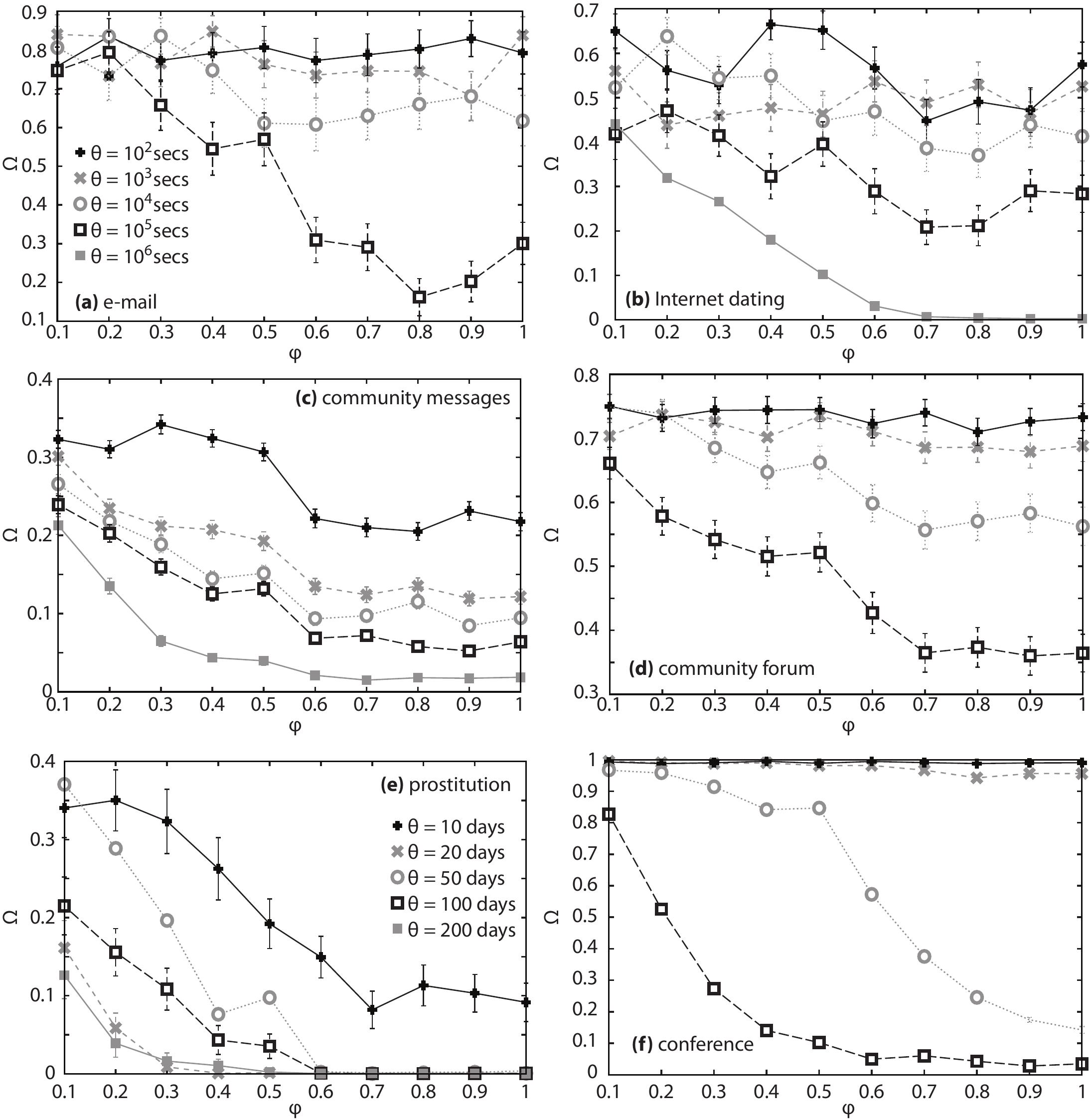}
  \caption{ Cascade size $\Omega$ versus threshold values $\phi$
from $0.1$ to $1.0$ for various time window sizes $\theta$. The error bars indicate the standard error over 200 runs of cascade simulations.}\label{fig:cascade-threshold}
\end{figure*}

The threshold is fixed and identical for all individuals. As the threshold increases, the agents are more resistance to change their initial states. In Watts's model, the cascade cannot occur if the  threshold value is too low~\cite{watts_threshold}. Introducing a time window  enables cascades to happen for higher threshold values. For short time windows only the most recent contacts counts. As larger time window integrate more influence, it can require more time to reach the threshold . It has been shown that in the static case, there is an upper limit for the threshold value beyond which no cascade can occur~\cite{watts_threshold}. Since the static case is equivalent to the temporal case when $\theta$ is equal to the duration of the data, the limiting threshold value will be at least as large in the temporal case. In Figure~\ref{fig:cascade-threshold}, we show the effect of increasing $\theta$ and $\phi$ on cascade size $\omega$ for six empirical datasets. In short time windows, individuals meet fewer people and meeting one adopter is enough to change the state. This behavior is similar to (hypothetical) disease spreading models with 100\% per-contact transmission probability.

\subsection{Effect of temporal structure on cascade size}
\label{subsec:2}

\begin{figure*}
  \includegraphics[width=1.0\linewidth]{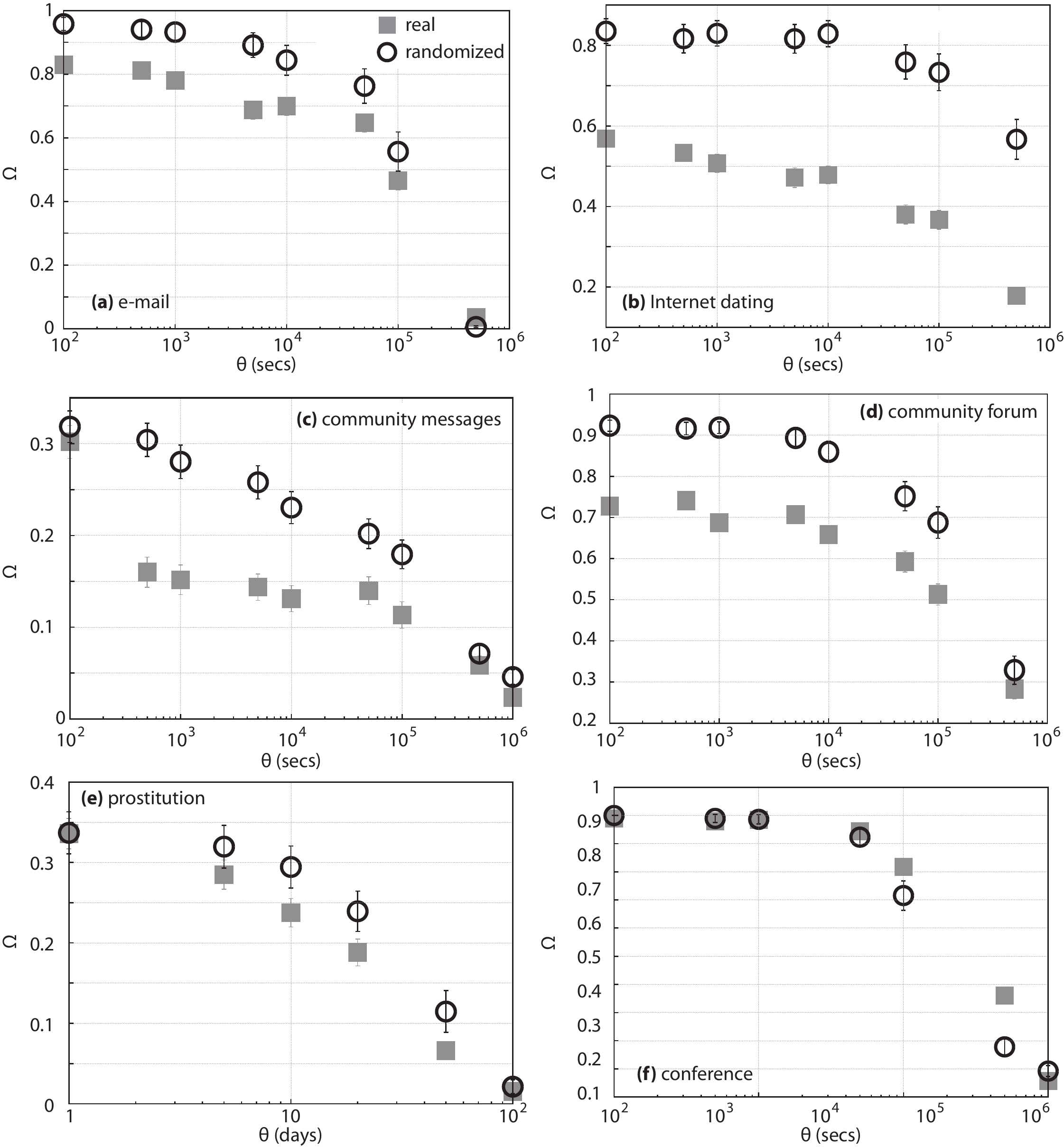}
  \caption{Cascade size versus time-window size for a fixed value of the threshold ($\phi=1/2$). To verify the effect of temporal correlations, we compare the data to a null model by reshuffling time stamps. The figures show that, except for the conference data, temporal correlations slow down the cascade size compare to the randomized networks without temporal structure. The error bars indicate the standard error over 150 runs of cascade simulations. The legend in panel (a) applies to all panels except (e).}\label{fig:Cascade-size-versus-time}
\end{figure*}

Now we will focus more on the effect of the temporal structure of contacts on the cascade size. Bursty behavior of contacts meaning that high intensity of contact activity in short interevent time and followed by no activity in long time intervals~\cite{barabasi_burst_2005,goh-burst}. Studies have shown the burstiness can influence spreading over the contacts~\cite{vazquez}. We examine the burstiness effect by applying our cascade model on the six empirical datasets and comparing the cascade size on the real and reshuffled time stamps. This reshuffled time stamps model (null model) follows Ref.~\cite{karsai_slow}. It preserves network structures such as degree distribution and the number of contacts. The null model is implemented by randomly reshuffling time stamps of contacts between pairs of nodes. Therefore burstiness and autocorrelations on single edges and pairwise correlations of edge pairs are destroyed. The null model, however, retains the global temporal statistics (number of events at given time, etc.~\cite{karsai_slow}).
 
Fig.~\ref{fig:Cascade-size-versus-time} represents the results of changing the size of the time window on cascade size for six empirical datasets. We choose the fractional threshold value to be $\phi=0.5$. Studies have been shown that in laboratory experiments, individuals' choice is well described by the threshold $0.5$~\cite{latane_threshold}.

The cascade size $\Omega$ decreases as a function of time window $\theta$. This is not surprising since a larger time window associate with more interaction per individuals and decreases the fraction of adopters one meets. For a very large time window, the cascade behavior is similar to the static threshold model---that cascade does not happen for larger threshold values~\cite{watts_threshold}. When $\theta$ is large enough, the temporal order of events no longer matters, exactly as we see in Figure \ref{eq:fractional_threshold} for large time windows.

For almost all datasets, the null model has a larger cascade size compare to the real datasets. In fact, other studies also suggest that the time correlation of events slows down the spreading~\cite{karsai_slow}. For conference data, the real and null models overlap for the small or large time windows. For intermediate-sized time windows, the real datasets have larger cascade sizes compare to the null model. We will try to explain this in terms of the network statistics. In the case of conference data, the network is densely connected (see Table~\ref{tab:properties-networks}). This high connectance guarantees that a sufficient fraction of adopters is exposed to individuals for the small and medium time windows. For the intermediate time windows, the temporal correlation plays an important role. This is especially strong in the conference data since it is organized around specific activities such as talks, poster sessions, coffee, and lunch breaks. Typically, networks of high assortativity have a densely connected subgraph---for the conference data, the entire network is such a densely connected graph.

\begin{table*}
\begin{tabular}{c|cccc}
The data & connectance & burstiness & clustering coefficient & assortativity\tabularnewline
\hline
Prostitution & 0.0002 & 0.44 & 0.00 [0.00] & $-0.10 [-0.02]$\tabularnewline
Email &0.006 & 0.68 & 0.06 [0.07] & $-0.25 [-0.22]$\tabularnewline
Conference  & 0.34 & 0.74 & 0.50 [0.48] & $-0.12 [-0.17]$\tabularnewline
Online dating & 0.0002 & 0.65 & 0.00 [0.00] & $-0.04 [-0.04]$\tabularnewline
Internet community forum & 0.012 &0.80 &0.43 [0.40]  & $-0.24 [-0.30]$ \tabularnewline
Internet community messages & 0.007 &0.65 &0.05 [0.08] &$-0.54 [-0.51]$  \tabularnewline
\end{tabular}
\caption{\label{tab:properties-networks}Properties of the aggregated (and hence static) networks. Connectance is the fraction of all possible links that are realized in a network~\cite{connectance}. Burstiness means a high activity in short inter-event time followed by long-range of no activity~\cite{goh-burst}. The clustering coefficient measures the fraction of triangles relative to the number of connected triples~\cite{newman2010}. Assortativity measures the correlation of the degrees at either side of an edge~\cite{newman2010}. The clustering coefficient and assortativity for null models are written in brackets.}
\end{table*}

\section{The time evolution of adoption in the fractional threshold model}

\begin{figure*}
  \includegraphics[width=1.0\linewidth]{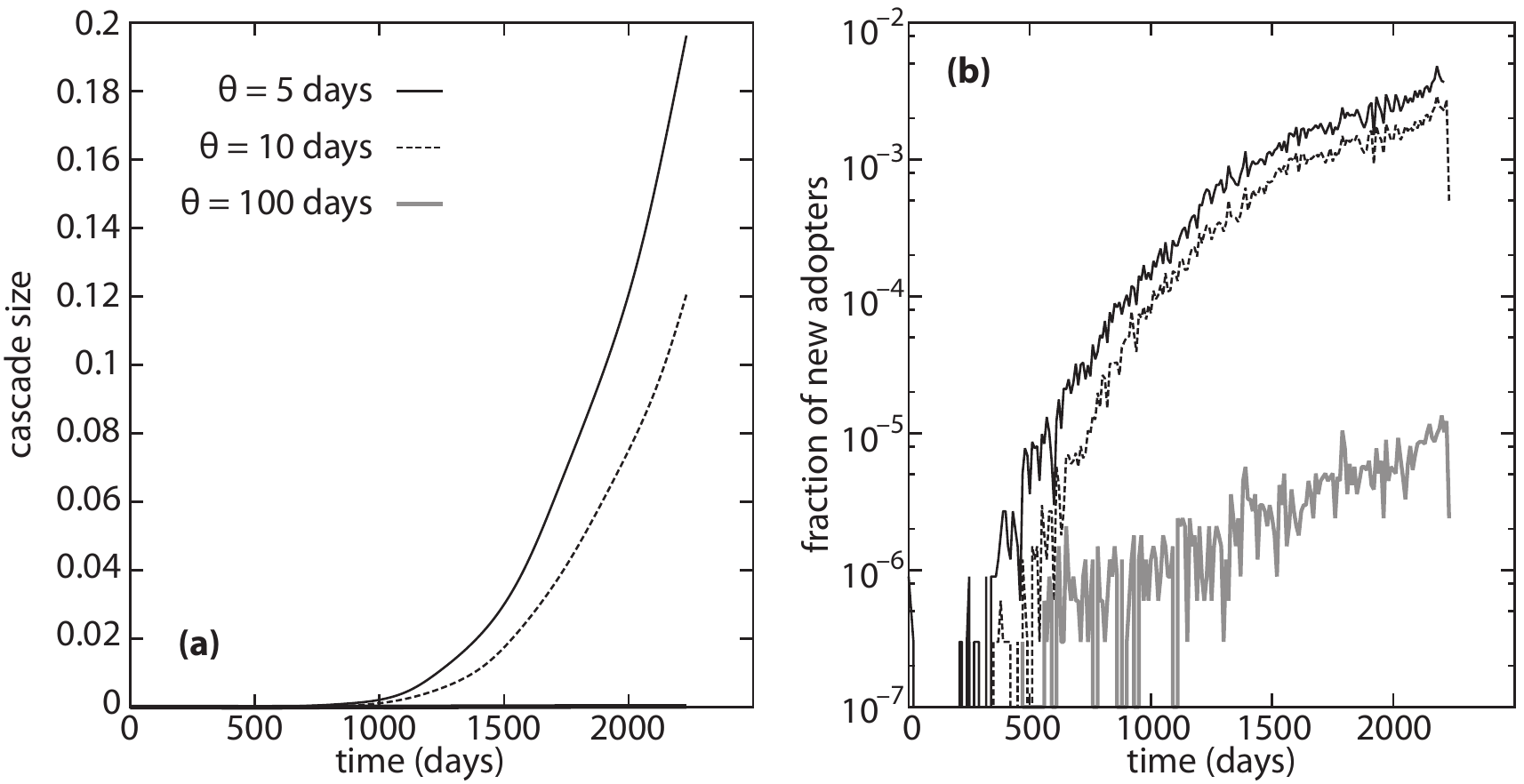}
  \caption{Evolution of cascade sizes versus sampling time
for the prostitution dataset. The threshold is $\phi=0.70$.}\label{fig:brazil_step_cas}
\end{figure*}

\begin{figure*}
  \includegraphics[width=1.0\linewidth]{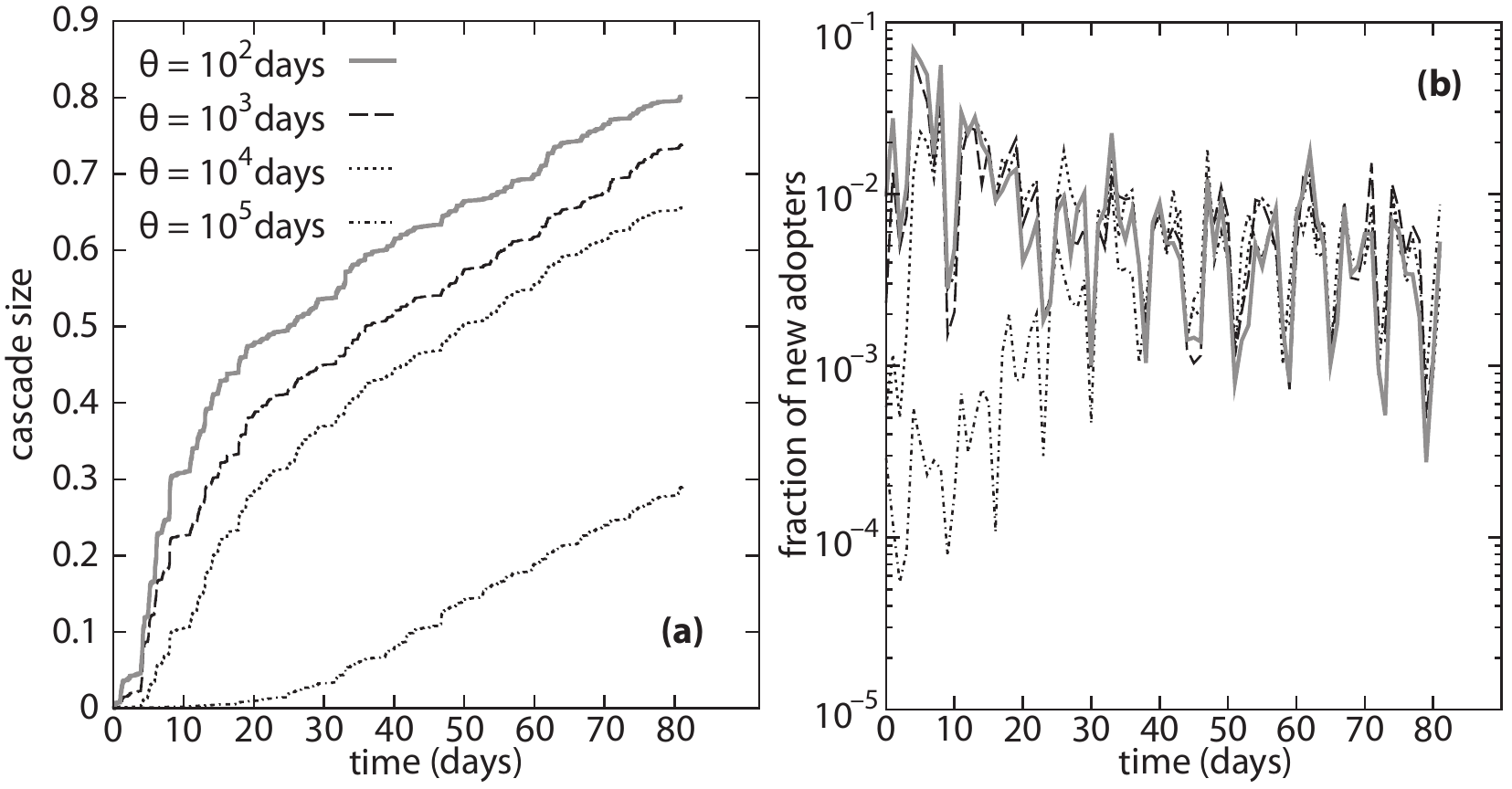}
  \caption{Evolution of cascade size versus sampling time
for the email dataset. The threshold is $\phi=0.70$.}\label{fig:eml2_step_cas}
\end{figure*}

\begin{figure*}
  \includegraphics[width=1.0\linewidth]{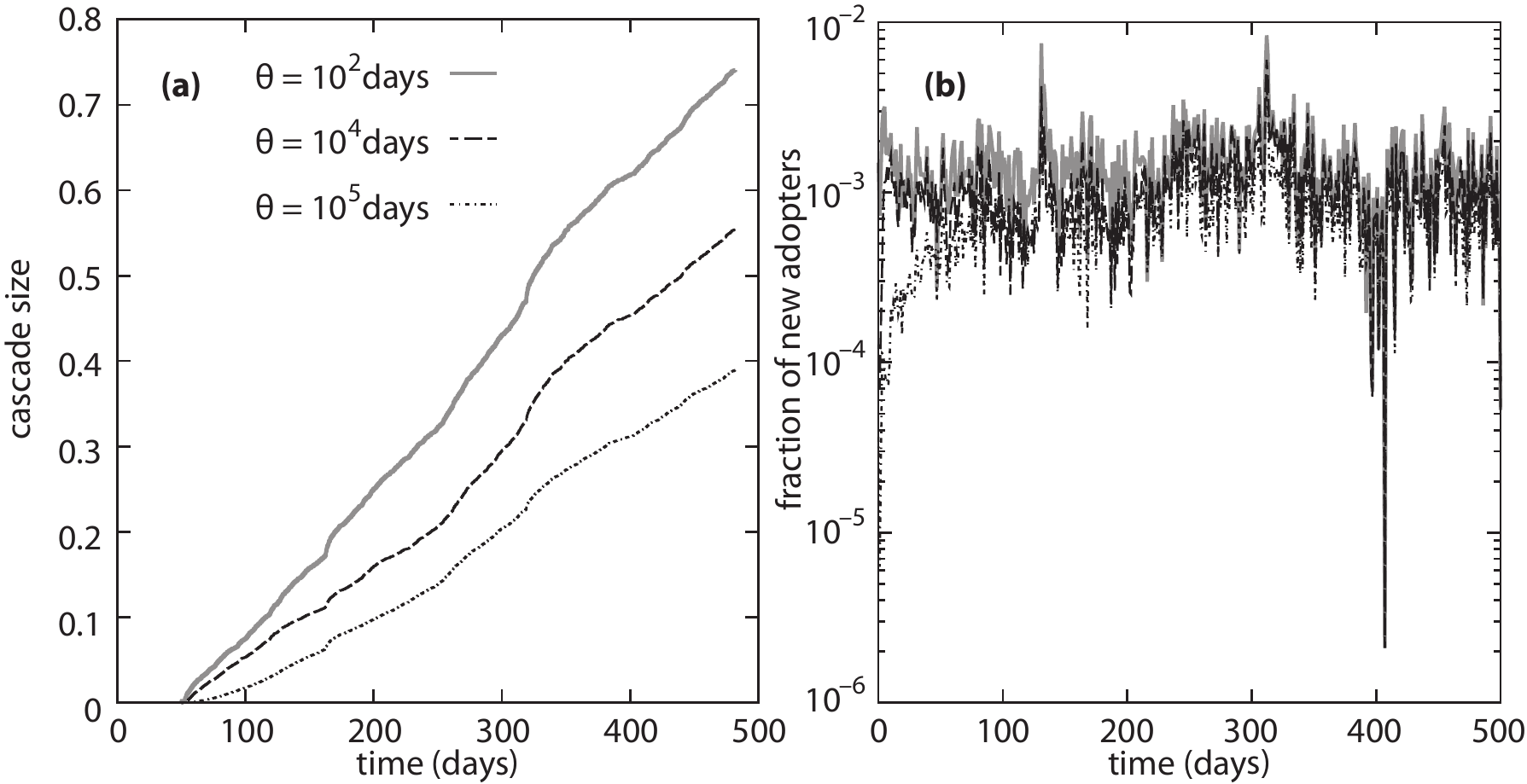}
  \caption{Evolution of cascade size versus sampling time
for the community forum dataset. The threshold is $\phi=0.70$.}\label{fig:ft_forum_short_step_cas}
\end{figure*}

\begin{figure*}
  \includegraphics[width=1.0\linewidth]{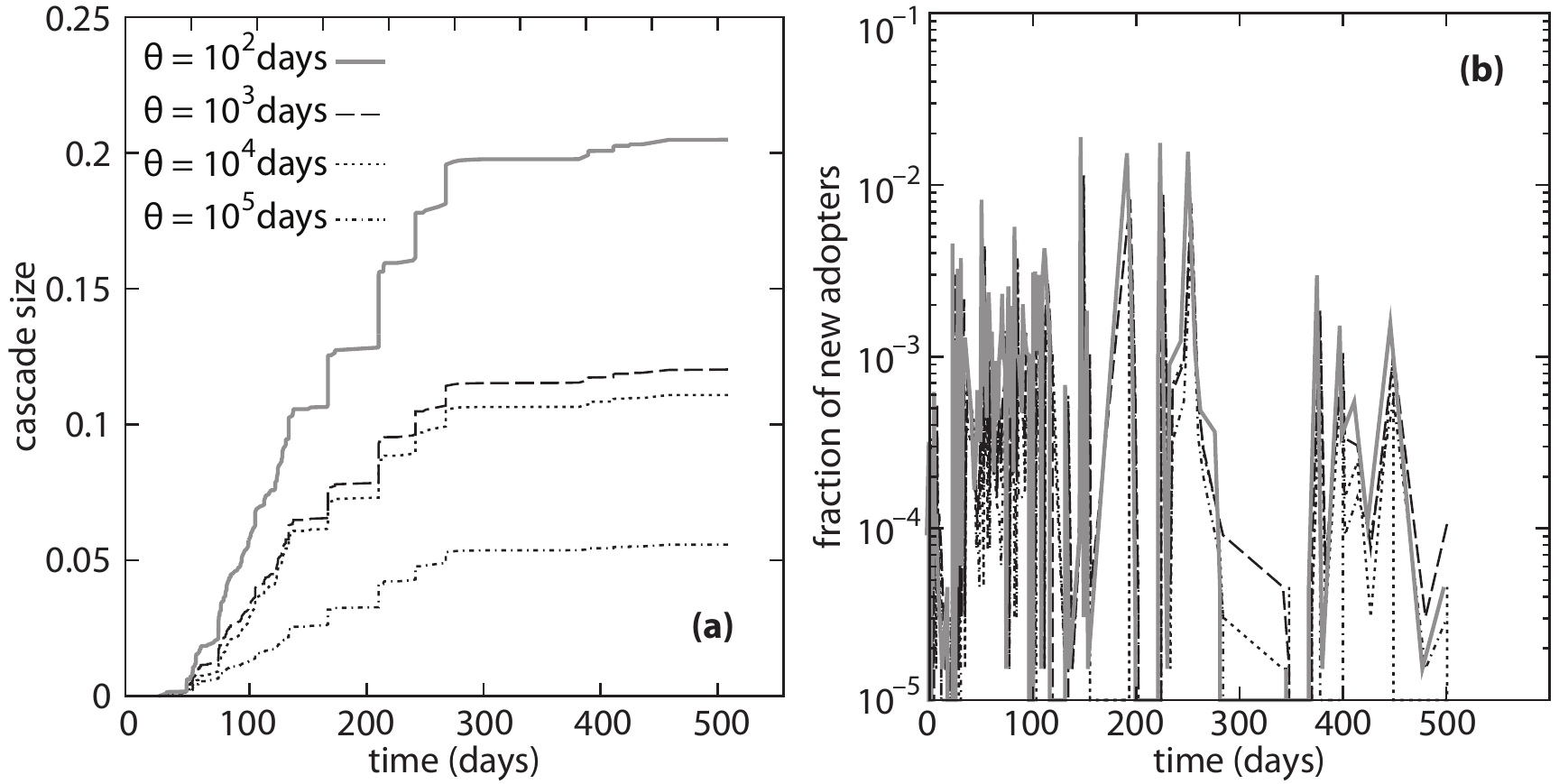}
  \caption{Evolution of cascade size versus sampling time
for the community messages dataset. The threshold is $\phi=0.70$.}\label{fig:ft_messages_short_step_cas}
\end{figure*}

\begin{figure*}
  \includegraphics[width=1.0\linewidth]{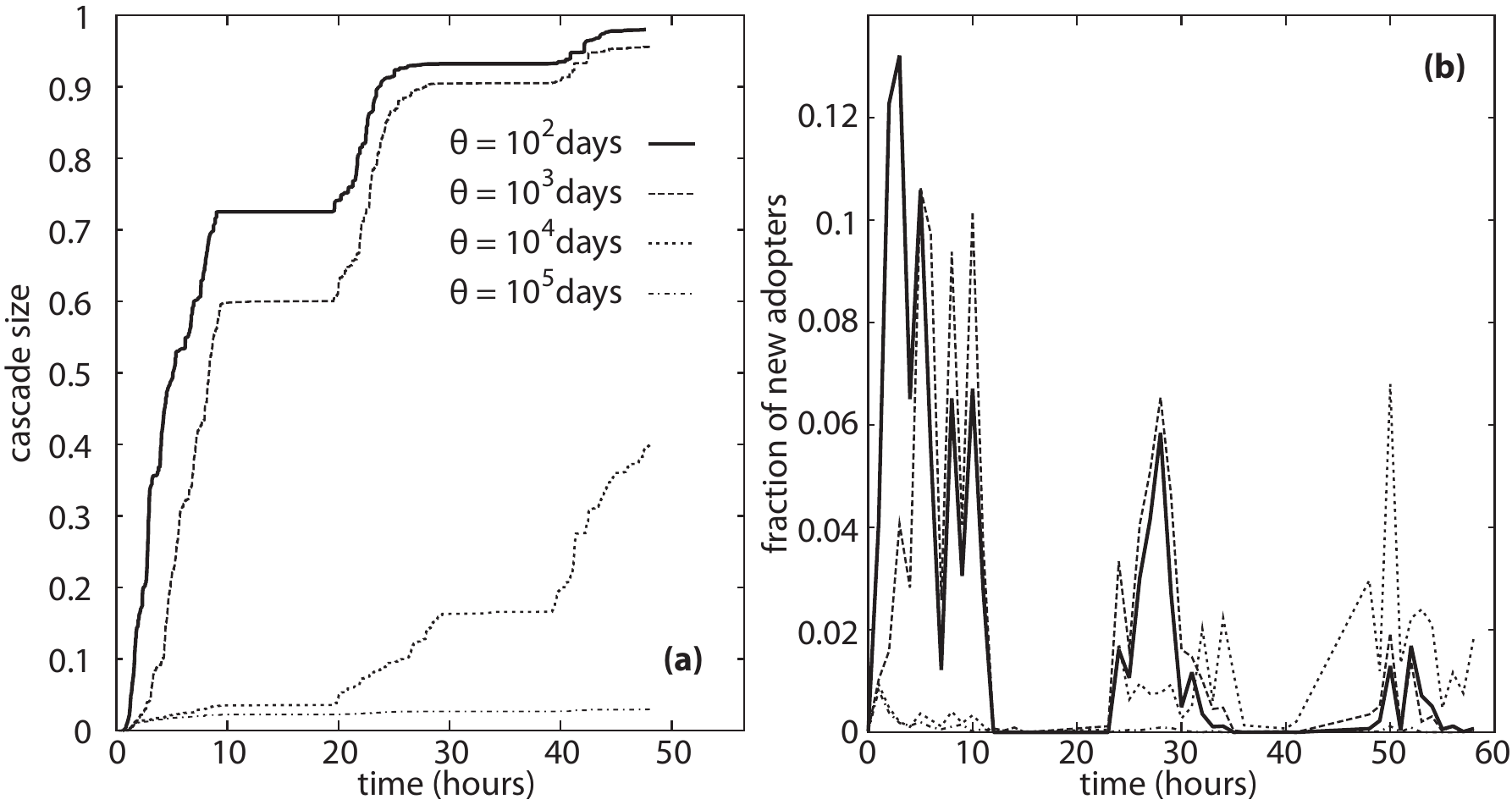}
  \caption{Evolution of cascade size versus sampling time
for the conference dataset. The threshold is $\phi=0.70$.}\label{fig:ht_contact_step_cas}
\end{figure*}

\begin{figure*}
  \includegraphics[width=1.0\linewidth]{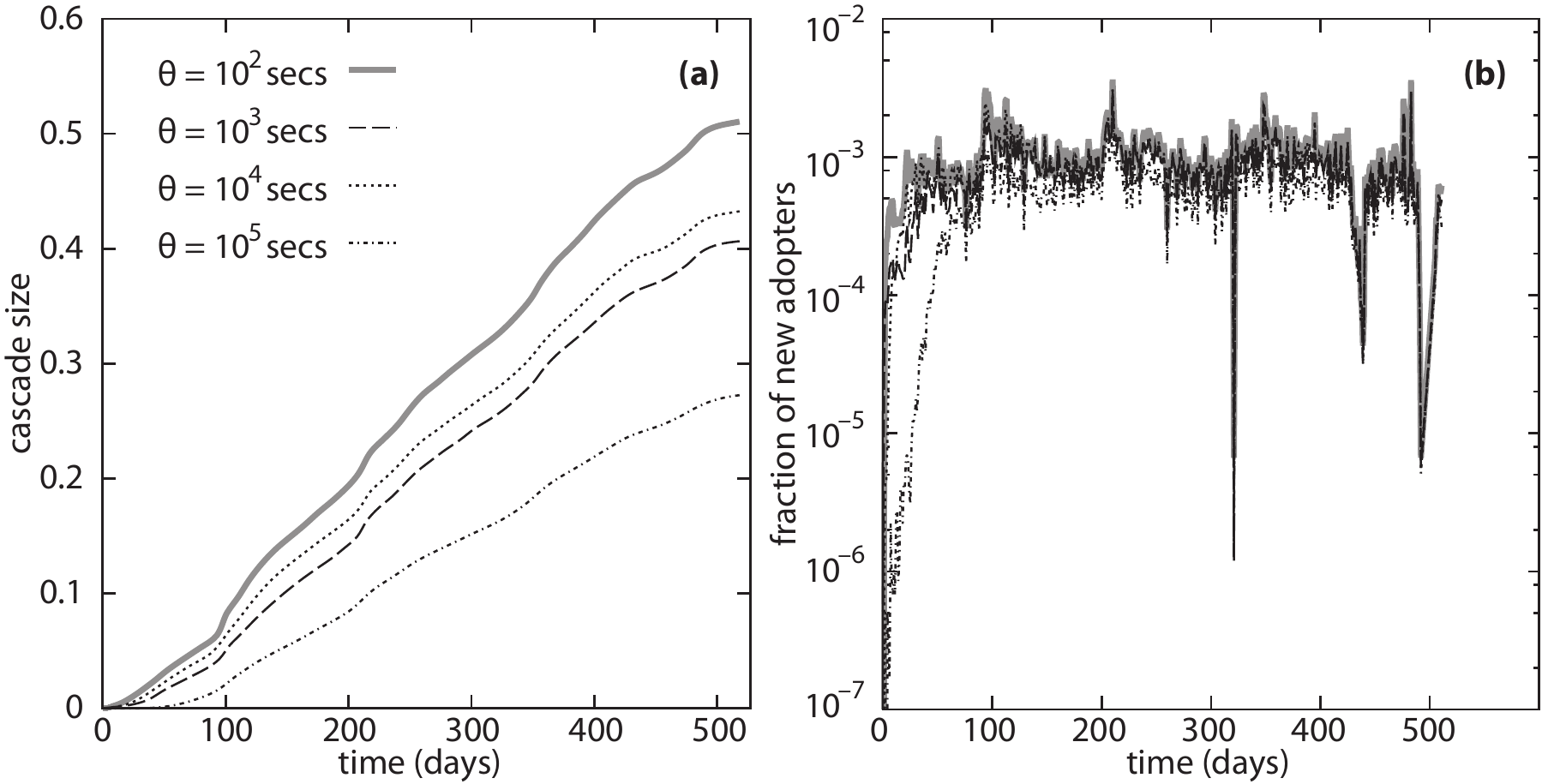}
  \caption{Evolution of cascade size versus sampling time
for the dating dataset. The threshold is $\phi=0.70$.}\label{fig:pok_step_cas}
\end{figure*}

Authors have studied spreading dynamics by the SI model of contagion and compared with null models~\cite{karsai_slow}. It has also been shown that heterogeneity in human activity affects the spreading~\cite{iribarren_email}. In this section, we study the spreading dynamics within a population on empirical datasets. Besides, we also study how the number of new adopters evolves in different time window sizes.  We investigate the time evolution of the cascades by two types of plots.  The first is the aggregated cascade size over time which has been defined in Eq.~\ref{fig:adopters}. The other plot is based on the number of new adopters for each time step in the evolving network. Figs.~\ref{fig:brazil_step_cas} through \ref{fig:pok_step_cas} show these two plots separately for each dataset.

The general pattern we observe here is that increasing the time window implies a decrease in the cascades' speed.  The larger time windows take more time for the vertex to change state. This effect is furthermore highly non-linear---sometimes the change is dramatic, sometimes hardly visible. The heterogeneity in the daily pattern is responsible for the jagged evolution of cascades. This effect is specifically clear in the curves of new adopters as a function of time. We also note that the growth is convex in some cases, concave in others (i.e., the cascade accelerates, like Fig.~\ref{fig:brazil_step_cas}, after a quiet start. This is presumably related to this dataset's sparseness and the fact that the overall activity in it grows with time. For Figs.~\ref{fig:ft_forum_short_step_cas} and \ref{fig:pok_step_cas} the outbreak progresses close to linearly. This is interesting in context of the Fig.~\ref{fig:ft_messages_short_step_cas} which represents the same set of users as  Fig.~\ref{fig:ft_forum_short_step_cas} but a different type of activity (forum posts rather than e-mail like messages). This could suggest that opinion cascades spread faster in open discussions than in person-to-person communication.

The datasets of Internet community messages and conference encounters are two examples of rather small networks. The distinctive difference is that the message data is sparser than the conference network. In the adoption dynamics, we see a similar pattern in both datasets---that as new adopters emerge over time, the activities speed up. The conference data's cascades' sizes seem to be much higher than the community message data, probably due to its higher contact density.

\section{Absolute threshold model}

The fractional threshold models explain only parts of socio-economical spreading processes. In reality, individuals may have a different threshold depending on whom they interact with. Even a single discussion with a relative or close friend can be sufficient for changing an opinion. In this chapter, we denote absolute threshold models, $\Phi$, the sufficient number of adopters the individual has to meet to change state, Eq.~\ref{eq:absolute_threshold}. One can assume that the reality is somewhere in between pure fractional or pure absolute threshold in many situations.

\begin{figure*}
  \includegraphics[width=1.0\linewidth]{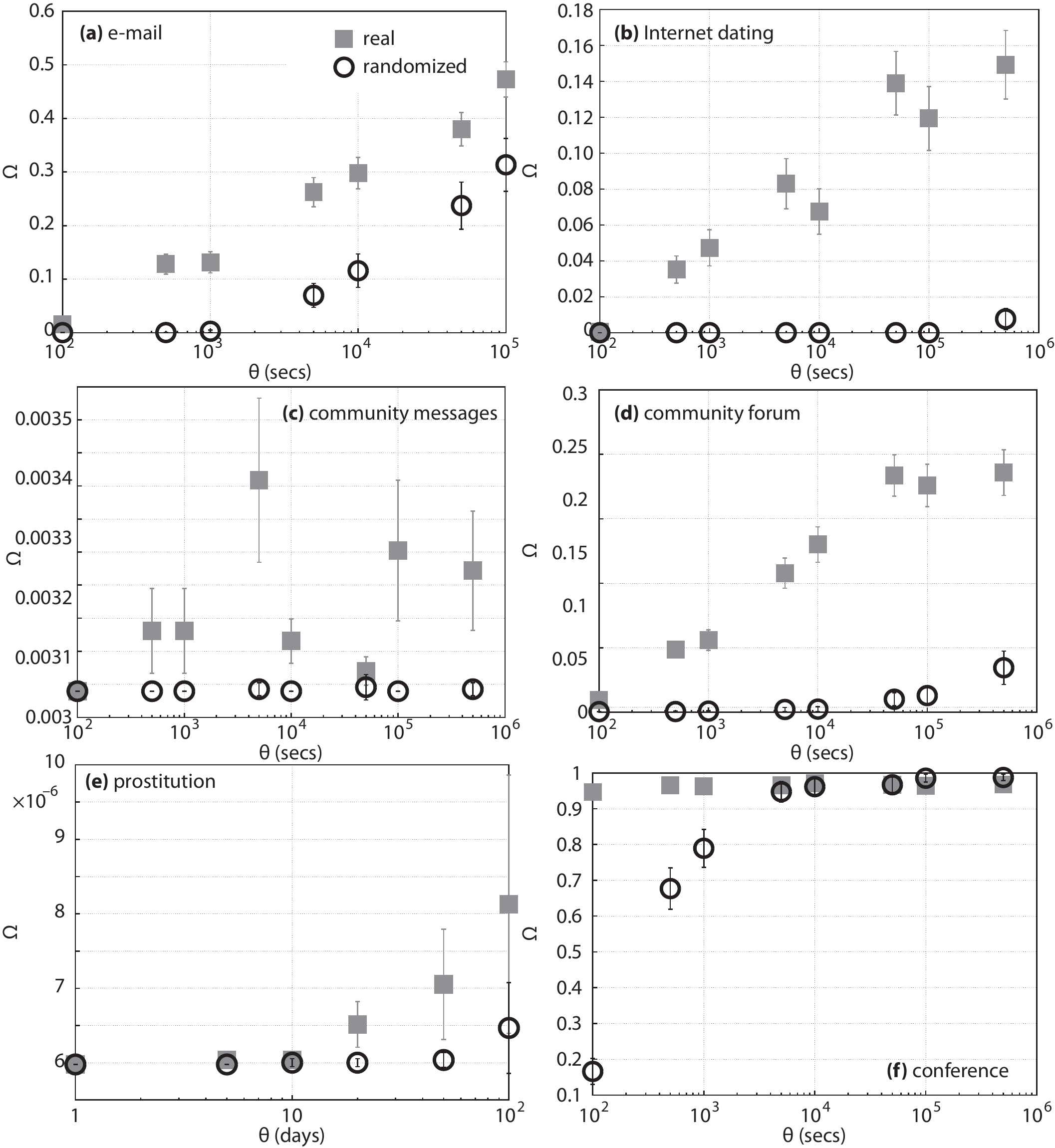}
  \caption{Cascade size versus time windows for the absolute-threshold model. The threshold is $\Phi=3$. To investigate the effect of temporal correlations, we compare the data to a null model where the times of contacts are randomly shuffled. In most cases, the temporal structure makes the cascades larger as the time windows increase. The error bars indicate the standard error over 200 runs of cascade simulations.}\label{fig:Cascade-size-absolute}
\end{figure*}

In Fig.~\ref{fig:Cascade-size-absolute} shows the cascade size as a function of various time windows. We fixed the absolute threshold value for all cases $\Phi=3$. As the figure shows, the cascade size increase by increasing the time window $\theta$. In a larger time window, the chance of meeting three adopters increases. Interestingly, for the time-reshuffled models, this is not the case. In fact, the time order in the empirical datasets boosts the cascade compare to randomized time order. In conference data, Fig.~\ref{fig:Cascade-size-absolute}(f), the sampling time is short. Therefore the time order does not play a major role. The increase in the cascade size by increasing the time window is the opposite of the fractional threshold models. Exposing to more contacts in the fractional model decrease the fraction of adopters in the time window. Moreover, opposite to the fractional threshold model, the absolute threshold model's null model has a smaller cascade size than empirical data. Thus as temporal correlation in fractional threshold model slows down the cascades. In the absolute threshold models, the opposite effect occurs. In the case of community messages---Fig.~\ref{fig:Cascade-size-absolute}(c)---since the number of contacts is low, there are large error bars that do not rule out an increasing $\theta$-dependence.

\section{Discussion}

We proposed an extension of Watts's cascade model~\cite{watts_threshold} to temporal networks where the influence is integrated over a sliding window of the contacts. This model builds on the key assumption that people are influenced by contacts dating for a certain duration into the past but not influenced by very old contacts. We studied two variants of the model, one where people respond to a threshold of the fraction of adopters among their neighbors and another where they respond to the absolute number of such contacts. We find that the dynamics of cascades are heavily affected by contacts' timing in the temporal networks.

There is a qualitative difference between the fractional- and absolute-threshold models. In the former case, the cascade sizes decrease with time-window size; in the latter case, it is the other way around. Moreover, the response to randomization is different---for the fractional-threshold case, the temporal-network structure makes the cascades larger. In contrast, in the absolute-threshold case, randomization decreases the size of cascades. This is interesting in the light of Ref.~\cite{karsai_slow} where the authors argue that burstiness slows down spreading phenomena. In this case, the authors have models of contagion in mind, but their conclusion seems not to generalize to threshold models (as was also observed in Refs.~\cite{takaguchi_etal} and \cite{karimi_holme}).

We also note that the cascades' time evolution differs from dataset to dataset, where the more sparsely connected systems show accelerating outbreaks. That explains why it takes a longer time for a cascade to take off, as there are some datasets where cascades fail to reach a large proportion of the population for a large part of parameter space. It is indeed hard to single out one temporal network structure that controls the cascade dynamics. But the average number of contacts per edge and the connectance (or average degree in the aggregated network) seem to be two important quantities.

We believe there are interesting open questions regarding threshold models in temporal networks; the main one is what structures---temporal or topological---control spreading phenomena in temporal networks.

\begin{acknowledgement}
The authors acknowledge financial support by the Swedish Research Council and the WCU
program through NRF Korea funded by MEST R31--2008--10029. The authors thank Taro Takaguchi for comments.
\end{acknowledgement}

\bibliographystyle{abbrv}
\bibliography{cascade_references}

\end{document}